\documentclass[12pt,aps,prd,nofootinbib,floats,floatfix,preprintnumbers,groupedaddress]{revtex4}

\usepackage{comment}
\usepackage{graphicx}
\usepackage{amsmath, mathrsfs}
\usepackage{float}
\usepackage{subfigure}
\usepackage[usenames]{color}
\usepackage{amssymb}
\usepackage{bbm}
\usepackage{csquotes}
\usepackage{color}
\usepackage{psfrag}
\usepackage{graphicx}
\usepackage{orcidlink}
\usepackage{caption,subcaption}
\usepackage{marginnote}
\usepackage{mathrsfs}
\usepackage[utf8]{inputenc}
\usepackage{tikz}
\usetikzlibrary{positioning,decorations.pathmorphing}

\usepackage[section]{placeins}
\usepackage{epstopdf}
\newcommand{\ta}{\tilde a}
\newcommand{\bea}{\begin{aligned}}
\newcommand{\eea}{\end{aligned}}
\newcommand{\beq}{\begin{equation}}
\newcommand{\eeq}{\end{equation}}

\newcommand{\bse}{\begin{subequations}}
\newcommand{\ese}{\end{subequations}}

\usepackage{adjustbox}

\usepackage{hyperref}
\hypersetup{
     colorlinks   = true,
     citecolor    = red,
     linkcolor    = blue,
     urlcolor     = blue,
}

\newcommand{\bmm}{\begin{multline}}
\newcommand{\emm}{\end{multline}}

\begin{document}

\title{\Large{Kerr-de Sitter Black Holes: Quantum Aspects and Cosmic Censorship Conjecture}}
\author{\bf{Ankit Anand}\orcidlink{0000-0002-8832-3212}}
\email{Anand@iitk.ac.in}
\affiliation{Department of Physics, Indian Institute of Technology Kanpur, Kanpur 208016, India.}
\author{\bf{Ankur Dey}\orcidlink{0009-0001-1077-0442}}
\email{ ankurd21@iitk.ac.in}
\affiliation{Department of Physics, Indian Institute of Technology Kanpur, Kanpur 208016, India.}

\pagenumbering{arabic}

\renewcommand{\thesection}{\arabic{section}}
\begin{abstract}
\begin{center}
    \hspace{1cm}
\end{center}

In this article, we test the validity of the weak cosmic censorship conjecture (wCCC) in the background of a quantum Kerr–de Sitter (qKdS) black hole, incorporating exact backreaction from quantum matter fields. Using a test particle approach, we analyze whether an extremal qKdS black hole can be over-extremized to expose a naked singularity. Our results indicate that the black hole remains stable against horizon-destroying processes induced by infalling matter. Quantum corrections enhance the horizon's robustness rather than destabilize it, offering further evidence that wCCC remains preserved in the presence of quantum effects.


\end{abstract}


\maketitle

\textbf{Keywords: }

\section{Introduction}

Gravitational singularities, first identified in the seminal works of Penrose and others~\cite{Penrose:1964wq, Brady:1995ni}, represent regions of spacetime where curvature invariants—such as the Kretschmann scalar—diverge. At such points, the classical description of general relativity breaks down and the Einstein field equations lose their predictive power. When these singularities are not concealed behind an event horizon, they become \textit{naked singularities}~\cite{Boulware:1973tlq}, allowing signals to escape from regions of arbitrarily large curvature. Their existence would undermine determinism, violate global hyperbolicity, and jeopardize the causal structure of spacetime.

To prevent such pathologies, Penrose proposed the Weak Cosmic Censorship Conjecture (wCCC)~\cite{Penrose:1969pc, Penrose:1964wq}, asserting that singularities formed through gravitational collapse are always hidden behind event horizons. The conjecture remains unproven, yet it plays a foundational role in classical and semiclassical gravity, underlying results such as the Penrose inequality~\cite{Penrose:1973um} and Hawking's area theorem~\cite{Hawking:1971tu}. A key approach to testing wCCC was introduced by Wald~\cite{Wald:1974hkz}, who considered whether an extremal Kerr--Newman black hole could be over-charged or over-spun by absorbing a test particle. The attempt failed: extremal solutions satisfy mass-dependent bounds~\cite{Newman:1965my} that prevent horizon destruction. Subsequent analyses incorporating self-force and backreaction effects~\cite{Mazur:1982db, Barausse:2010ka, Sorce:2017dst} further demonstrated that classical black holes are generically protected from over-extremization.

Beyond classical general relativity, a wide range of attempts to violate wCCC have been explored through quantum tunneling~\cite{Matsas:2007bj, Richartz:2008xm, Bambi:2011yz, Yang:2022yvq}, test fields~\cite{Duztas:2013wua}, thin shells~\cite{Mann:2008rx}, superradiant scattering~\cite{Semiz:2015pna}, and numerical collapse~\cite{Choptuik:1992jv, Joshi:1993zg}. Although higher-dimensional spacetimes can exhibit long-wavelength instabilities in black strings, rings, or ultra-spinning holes leading to short-lived naked singularities~\cite{Lehner:2010pn, Figueras:2017zwa}, these violations are generally weak or non-generic~\cite{Emparan:2020vyf}. In four dimensions, robust violations remain elusive. Nevertheless, recent semiclassical studies indicate that quantum corrections may destabilize horizons under certain circumstances. For example, Ref.~\cite{Li:2013sea} showed that semiclassical effects could eliminate the horizon of a regular black hole, suggesting that quantum processes may circumvent the classical protection mechanism assumed in wCCC.

These developments motivate examining wCCC within quantum-corrected geometries. In this work, we investigate the conjecture in the context of the quantum-corrected Kerr–de Sitter (qKdS) black hole introduced in Ref.~\cite{Emparan:2020znc}. This solution arises from a braneworld realization of a rotating dS$_4$ C-metric and incorporates full backreaction from quantum fields. On the brane, it generalizes the Kerr–de Sitter (KdS) solution~\cite{Banados:1992wn, Banados:1992gq} by introducing a parameter $\gamma$ that encodes the strength of the quantum backreaction. The existence of the event horizon imposes a nontrivial inequality relating the mass, angular momentum, and $\gamma$. Saturation of this relation defines an extremal configuration analogous to extremal Kerr--Newman black holes.

We revisit Wald's thought experiment in this quantum-corrected background, analyzing whether an infalling test particle can drive an extremal qKdS black hole beyond its extremality bound and thereby expose a naked singularity. Our analysis is carried out perturbatively, tracking the first-order changes in the conserved quantities induced by particle absorption and testing whether the final configuration satisfies or violates the horizon condition. This allows us to determine whether quantum backreaction weakens or reinforces the protective mechanism of cosmic censorship.

The present work provides the first systematic test of the Weak Cosmic Censorship Conjecture in a rotating de Sitter geometry that incorporates exact quantum backreaction from strongly coupled conformal fields. While related analyses have been performed for the quantum BTZ black hole and for semiclassical regular black holes, no analogous study has been carried out for the fully backreacted quantum Kerr–de Sitter \(qKdS_3\) solution. Our approach extends Wald’s gedanken experiment to this quantum-corrected background, and, crucially, includes both first- and second-order perturbative corrections in the spirit of the Sorce–Wald formalism. This establishes the \(qKdS_3\) spacetime as a controlled laboratory where the interplay between rotation, de Sitter curvature, and quantum backreaction can be examined with precision.

This article is organized as follows. In Sec.~\ref{Sec:qKdS Black hole}, we review the qKdS geometry and highlight the conditions governing extremality. Sec.~\ref{Sec:Test of wCCC} presents the dynamics of test particles and the perturbative analysis required to probe wCCC in this background. We conclude in Sec.~\ref{sec:conclusions} with a summary of our findings and their implications for quantum-corrected black hole spacetimes.

\section{Review of the Quantum Kerr--de Sitter Black Hole}\label{Sec:qKdS Black hole}

Brane-localized black hole solutions, first introduced in \cite{Emparan:1999fd}, represent a major development in the study of higher-dimensional and braneworld gravity. A subsequent interpretation based on the AdS/CFT correspondence was proposed in \cite{Emparan:2002px}, where the solution was identified as a quantum-corrected BTZ black hole, incorporating the full semi-classical backreaction generated by conformal fields living on the brane. This programme was later refined in \cite{Emparan:2020znc}, where the effective lower-dimensional field equations and their exact backreacted solutions were derived. The inclusion of bulk rotation and the resulting brane-induced angular momentum were discussed in detail in \cite{Panella:2023lsi}, providing the foundation for the rotating quantum Kerr--de Sitter (qKdS) geometry. These developments have inspired continued interest in quantum black holes in reduced-dimensional or braneworld settings \cite{Emparan:2022ijy, Panella:2023lsi} and have clarified several aspects of their thermodynamic behaviour \cite{Frassino:2022zaz}.

Three-dimensional quantum black holes offer a uniquely tractable framework for probing the fundamental mechanisms that control horizon stability. Unlike four-dimensional semiclassical spacetimes, where backreaction typically requires perturbative or numerical treatment, the $qKdS_3$ geometry incorporates all-orders quantum corrections in closed analytic form. This makes it an ideal setting for conceptual studies such as cosmic censorship, holographic thermodynamics, and extremality bounds. Moreover, the brane-world origin of the $qKdS_3$ solution connects its dynamics directly to the holographic CFT living on the brane, enabling a clean interpretation of quantum stress-tensor effects on black hole horizons. These features render the $qKdS_3$ family a powerful model for disentangling quantum and geometric contributions to horizon protection mechanisms

\subsection{Metric and Horizons}

We now summarize the properties of the quantum Kerr--de Sitter black hole. The starting point is the Kerr--AdS$_4$ C-metric, an exact type-D solution belonging to the Plebański--Demiański family \cite{Plebanski:1976gy}. In the coordinate system convenient for braneworld embedding, the bulk metric takes the form
\begin{eqnarray}\label{Bulk Metric}
   ds^{2}=\frac{\ell^2}{(\ell + yr)^2}\biggr(\!-\frac{\Tilde{\mathcal{F}}(r)}{\Upsilon}(dt-ay^{2}d\varphi)^{2}
   +\frac{\Upsilon}{\Tilde{\mathcal{F}}(r)}dr^{2}
   +r^{2}\Big[\frac{\Upsilon}{\mathcal{J}(y)}dy^{2}
   +\frac{\mathcal{J}(y)}{\Upsilon}\Big(d\varphi+\frac{a}{r^{2}}dt\Big)^{2}\Big]\biggr)
\end{eqnarray}
with metric functions
\begin{equation}
\Tilde{\mathcal{F}}(r)=1-\frac{r^{2}}{\ell_{3}^{2}}-\frac{\mu\ell}{r}+\frac{a^{2}}{r^{2}}, \qquad
\mathcal{J}(y)=1-y^{2}-\mu y^{3}-\frac{a^{2}}{\ell_{3}^{2}}y^{4}, \qquad
\Upsilon=1+\frac{a^{2}y^{2}}{r^{2}} .
\end{equation}
Here, $\ell$ is the acceleration parameter, $\mu$ encodes the mass deformation, and $\ell_{3}$ is the induced de Sitter radius on the brane. These parameters satisfy
\begin{equation}
    \ell = \frac{\ell_3 \ell_4}{\sqrt{\ell_3 + \ell_4^2}},
\end{equation}
while the rotation parameter $a$ governs the bulk angular momentum. Regularity of the angular sector requires $\mathcal{J}(y)$ to admit at least one positive root; we denote its smallest positive root by $y_1$. The hypersurface $y=0$ inherits a well-defined geometry and naturally serves as the location of the brane. The induced three-dimensional metric is then
\begin{equation}\label{Brane Metric in tr}
  ds^2\big|_{y=0} = -\Tilde{\mathcal{F}}(r)\,dt^2 
  + \frac{dr^2}{\Tilde{\mathcal{F}}(r)} 
  + r^{2}\Big(d\varphi + \frac{a}{r^{2}}dt\Big)^{2} \ .
\end{equation}

To remove the angular conical singularity and restore the correct periodicity of $\varphi$, the coordinate redefinition introduced in \cite{Panella:2023lsi} is employed. Defining
\begin{equation}
\Delta = \frac{2 y_{1}}{3-y_{1}^{2}+\tilde{a}^{2}}, 
\qquad 
\tilde{a}\equiv\frac{ay_{1}^{2}}{\ell_{3}},
\end{equation}
and performing the transformations
\begin{equation}
 t = \tilde{t} + a y_1^{2}\tilde{\varphi}, \qquad \varphi = \tilde{\varphi}, \qquad
 \tilde{t} = \Delta \bar{t}, \qquad \tilde{\varphi} = \Delta \bar{\varphi},
\end{equation}
one obtains a globally well-defined rotating geometry in the barred coordinates $(\bar{t},\bar{r},\bar{\varphi})$:
\begin{eqnarray}\label{qKdS Metric}
    ds^{2}_{\text{qKdS}} &=& -g_{\bar{t}\bar{t}} \, d\bar{t}^{2} 
    + g_{\bar{\phi}\bar{\phi}} \, d\bar{\varphi}^{2}  
    + g_{\bar{r}\bar{r}} \, d\bar{r}^{2}  
    - g_{\bar{\phi}\bar{t}} \, d\bar{\varphi}\,d\bar{t} \nonumber\\
&=& -\left(1 - 8\mathcal{G}_{3}M - \frac{\bar{r}^{2}}{\ell_3^{2}} - \frac{\mu\ell\Delta^{2}}{r}\right) d\bar{t}^{2}
   +\left( \bar{r}^{2} + \Psi \frac{ \tilde{a}^2\,\bar{r}^2}{(1+\tilde{a})^2\Delta^2 r^2} \right) d\bar{\varphi}^{2}  \\
&&\quad \underbrace{\left[\left(1 - 8\mathcal{G}_{3}M - \frac{\bar{r}^{2}}{\ell_3^{2}} + \frac{(4\mathcal{G}_{3}J)^{2}}{\bar{r}^{2}} - \Psi\right)^{-1}\right]}_{\stackrel{}{\mbox{$\mathcal{F}(r)$}}}\! d\bar{r}^{2}
   - 8\mathcal{G}_{3}J\left(1 + \frac{\ell}{y_{1}r}\right)d\bar{\varphi}\,d\bar{t}. \nonumber
\end{eqnarray}
The functions $M$, $J$, and the quantum backreaction term $\Psi(\bar{r},y_1,\tilde{a})$ are given by
\begin{equation}
   \mu=\frac{1-y_{1}^{2}-\tilde{a}^{2}}{y_{1}^{3}}, \qquad
   \Psi = \frac{\mu\ell(1+\tilde{a}^{2})^{3/2}\Delta^{3}\sqrt{\bar{r}^2-r_S^2}}{\bar{r}^2},
\end{equation}
with
\begin{equation}
r_S = -\frac{2\tilde{a}\ell_3\sqrt{2-y_1^2}}{3-y_1^2+\tilde{a}^2},
\end{equation}
and
\begin{eqnarray}\label{M and J relation}
8\mathcal{G}_{3} M  = 1 - \frac{4\left[y_{1}^{2} - \tilde{a}^{2}(y_{1}^{2} - 4)\right]}{(3-y_{1}^{2}+\tilde{a}^{2})^{2}}, 
\qquad 
4\mathcal{G}_{3}J = \frac{4\ell_{3}\tilde{a}(y_{1}^{2}+\tilde{a}^{2}-1)}{(3-y_{1}^{2}+\tilde{a}^{2})^{2}}.
\end{eqnarray}
Here $\mathcal{G}_{3}$ denotes the renormalized three-dimensional Newton constant, $\mathcal{G}_{3}=L_{4}G_{3}/\ell$. The geometry \eqref{qKdS Metric} satisfies the full semi-classical field equations on the brane and is interpreted as the quantum Kerr--de Sitter black hole: a rotating $dS_3$ black hole sourced by the all-order backreaction of the brane CFT.

\subsection{Horizon Structure}

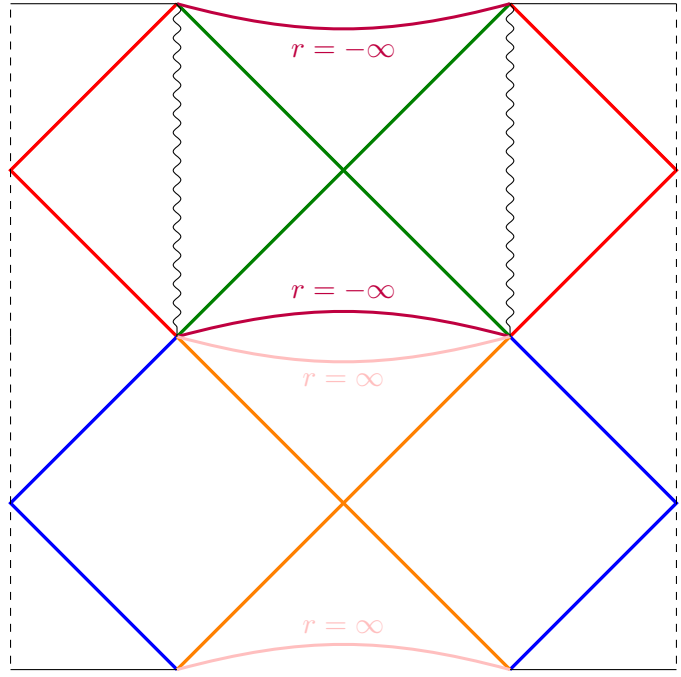
\begin{figure}[t!]
\centering
\begin{tikzpicture}[scale=1.1]
    \pgfmathsetmacro\myunit{4} 
    \draw [dashed, white] (0,0) coordinate (a) --++(90:\myunit) coordinate (b);
    \draw [white] (b) --++(0:\myunit) coordinate (c);
    \draw[dashed, white] (c) --++(-90:\myunit) coordinate (d);
    
    \coordinate (n3) at (0,-4);
    \coordinate (n4) at (4,-4);  

    \draw [line width = .4mm, red] (b) -- node[pos=.5, above, sloped]{${\color{red}}$} (-2,2) -- (a);
    \draw [line width = .4mm, red] (a) -- node[pos=.5, below, sloped] {${\color{red} }$} (-2,2) -- (b);
    \draw [line width = .4mm, red] (c) -- node[pos=.5, above, sloped] {${\color{red} }$} (6,2) -- (d);
    \draw [line width = .4mm, red] (d) -- node[pos=.5, below, sloped] {${\color{red} }$} (6,2) -- (c);

    \draw [line width = .4mm, blue] (a) -- node[pos=.5, above, sloped] {${\color{blue} }$} (-2,-2) -- (n3);
    \draw [line width = .4mm, blue] (n3) -- node[pos=.5, below, sloped] {${\color{blue} }$} (-2,-2) -- (a);
    \draw [line width = .4mm, blue] (n4) -- node[pos=.5, below, sloped] {${\color{blue} }$} (6,-2) -- (d);
    \draw [line width = .4mm, blue] (d) -- node[pos=.5, above, sloped] {${\color{blue} }$} (6,-2) -- (n4);

    \draw [line width = .4mm, green!50!black] (b) -- node[pos=.5, above, sloped] {${\color{green!50!black} }$} (2,2) -- (d);
    \draw [line width = .4mm, green!50!black] (d) -- node[pos=.5, above, sloped] {${\color{green!50!black} }$} (2,2) -- (b);
    \draw [line width = .4mm, green!50!black] (c) -- node[pos=.5, above, sloped] {${\color{green!50!black} }$} (2,2) -- (a);
    \draw [line width = .4mm, green!50!black] (a) -- node[pos=.5, above, sloped] {${\color{green!50!black} }$} (2,2) -- (c);

    \draw [line width = .4mm, orange] (a) -- node[pos=.5, above, sloped] {${\color{orange} }$} (2,-2) -- (n4);
    \draw [line width = .4mm, orange] (n4) -- node[pos=.5, above, sloped] {${\color{orange} }$} (2,-2) -- (a);
    \draw [line width = .4mm, orange] (n3) -- node[pos=.5, above, sloped] {${\color{orange} }$} (2,-2) -- (d);
    \draw [line width = .4mm, orange] (d) -- node[pos=.5, above, sloped] {${\color{orange} }$} (2,-2) -- (n3);

    \draw[dashed] (-2,0) coordinate (e) -- (-2,-4) coordinate (n1);
    \draw [line width = .4mm, purple] (a) to [out=15, in=165] node[pos=.5, above] {$r=-\infty$} (d);
    \draw [line width = .4mm, pink] (a) to [out=-15, in=-165] node[pos=.5, below] {$r=\infty$} (d);
    \draw [dashed] (e) -- (-2,4) coordinate (f);
    \draw (f) -- (b);
    \draw [line width = .4mm, purple] (b) to [out=-15, in=-165] node[pos=.5, below] {$r=-\infty$} (c);
    \draw (c) -- (6,4) coordinate (g);
    \draw [dashed] (g) -- (6,0) coordinate (h);
    \draw[dashed] (h) -- (6,-4) coordinate (n2);
    \draw [decorate, decoration={snake, amplitude=0.5mm, segment length=2.5mm}] (c) -- (d) node[pos=.5, above, sloped] {};
    \draw [decorate, decoration={snake, amplitude=0.5mm, segment length=2.5mm}] (b) -- (a) node[pos=.5, above, sloped] {};
    \draw (n1) -- (n3);
    \draw (n2) -- (n4);  
    \draw [line width = .4mm, pink] (n3) to [out=15, in=165] node[pos=.5, above] {$r=\infty$} (n4);
\end{tikzpicture}
\caption{\small Penrose diagram of a neutral quantum Kerr black hole in three-dimensional de Sitter spacetime (dS$_3$). The hypersurfaces $r = r_{-}$, $r = r_{+}$, $r = r_{c}$, and $r = r_{H}+r_C+r_{C_H}$ are depicted in red, blue, orange, and green, respectively, and the curvy line represents the singularity $r=0$. The diagram illustrates the global causal structure with periodic identifications along surfaces of constant $\bar{t}$. The vertical direction extends infinitely, while the dashed boundaries represent identified edges.}
\label{fig:penKdS}
\end{figure}

The horizon locations correspond to the real, positive roots of $g_{\bar{r}\bar{r}}^{-1}(\bar{r})$. For the qKdS solution one may write \cite{Panella:2023lsi}
\begin{equation}
 \frac{1}{ g_{{\bar{r}\bar{r}}}(\bar{r})} = 
 \frac{(r_C-\bar{r})(\bar{r}-r_H)(\bar{r}-r_{C_H})(\bar{r}+r_H+r_C+r_{C_H})}{\bar{r}^{2}\ell_{3}^{2}},
\end{equation}
where $r_H$, $r_{C_H}$, and $r_C$ denote, respectively, the event, Cauchy, and cosmological horizons. The remaining root is negative and lies behind the physical singularity at $\bar{r}=0$. For non-degenerate, real roots, the qKdS spacetime admits a maximal analytic extension following the standard approach of \cite{Gibbons:1977mu}. The complete Penrose diagram reveals an infinite temporal tower of regions, similar to the 4D Kerr--de Sitter case, though the presence of rotation generically induces closed timelike curves (CTCs). As discussed in \cite{Booth:1998gf}, a global periodic identification along constant-$\bar{t}$ hypersurfaces removes CTCs and consistently connects the two oppositely rotating sectors at the coincidence of the event and cosmological horizons ($r_H=r_C$). The resulting causal structure is shown in Fig.~\ref{fig:penKdS}.

In the static case, the horizons coincide with the Killing horizons of the timelike vector \(\partial_{t}\). Once rotation is switched on, however, the relevant null generator is promoted to
\[
\zeta^{b}=\partial_{t}-\frac{a}{r_{i}^{2}},\partial_{\phi}, ,
\]
which becomes null precisely at the horizon radii \(r_{i}\). The corresponding angular velocity of each rotating Killing horizon then follows as
\[
\Omega_{i}
= \frac{a}{\ell_{3}^{2}},\frac{y_{1}^{2} r_{i}^{2}-\ell_{3}^{2}}
{r_{i}^{2}+\tilde{a}^{2}y_{1}^{2}} , .
\]

The surface gravity \(\kappa_{i}\), defined through the standard relation
\[
\bar{\zeta}^{b}\nabla_{b}\bar{\zeta}^{c}= \kappa_{i},\bar{\zeta}^{c} \ ,
\]
takes the compact form
\[
\kappa_{i}=
\frac{\eta(1+\tilde{a}^{2})}{r_{i}^{2}+\tilde{a}^{2}y_{1}^{2}}
,\frac{1}{2\ell_{3}^{2}r_{i}},
\Big|\ell_{3}^{2}\mu\ell, r_{i}-2r_{i}^{4}-2\tilde{a}^{2}\ell_{3}^{2}\Big| .
\]

A striking feature of this expression is that the cosmological surface gravity \(\kappa_{c}\) vanishes exactly when the cosmological horizon merges with either the outer or inner quantum horizon, \(r_{c}=r_{+}\) or \(r_{c}=r_{-}\). Analogous degeneracy conditions hold for \(\kappa_{+}\) and \(\kappa_{-}\), signalling the onset of extremal configurations within the quantum Kerr–de Sitter family. In the non-rotating limit, our expressions smoothly collapse to the quantum-corrected Schwarzschild-de Sitter geometry of~\cite{Emparan:2022ijy}, while in the vanishing-backreaction regime \(r_{\pm}\rightarrow 0\) both \(\kappa_{+}\) and \(\kappa_{-}\) tend to zero. The associated Hawking temperatures follow the standard relation \(T_{i}=\kappa_{i}/(2\pi)\).

Depending on the relative placement and mutual degeneracies of the horizons, the spacetime displays a rich set of limiting phases, each capturing a distinct facet of the rotating quantum Kerr–de Sitter geometry.

\subsubsection*{Extremal Case}

An extremal black hole configuration arises when the outer event horizon $r_{H}$ and the inner Cauchy horizon $r_{C_H}$ merge into a single, degenerate root. At this point the surface gravity vanishes,
\begin{equation}
    \kappa_{H}=0 \,, \qquad T_{H}=0 \,,
\end{equation}
indicating a zero-temperature (``cold'') black hole. The extremal limit induces a qualitative modification of the global geometry: the degenerate horizon becomes infinitely distant, in proper length, from any other region of the spacetime. Consequently, the interior becomes causally disconnected from the asymptotic domain, a behaviour typical of extremal rotating or charged black holes.

\subsubsection*{Nariai Case}

The Nariai configuration corresponds to the critical situation in which the cosmological horizon $r_{C}$ coincides with the outer event horizon $r_{H}$,
\begin{equation}
    r_{C} = r_{H} \equiv r_{N} \, ,
\end{equation}
and is associated with a maximal mass parameter $\mu = \mu_{N}$. At this point the spacetime undergoes a geometric transition, and the near-horizon region factorizes as
\begin{equation}
    \mathrm{dS}_{2} \times S^{1} ,
\end{equation}
(or $\mathrm{dS}_{2} \times S^{d-2}$ in higher dimensions), characteristic of the Nariai limit.

For the rotating quantum Kerr--dS$_3$ background the Nariai radius is larger than in the static case. This behaviour mirrors that of the charged Nariai solution, suggesting that angular momentum and electric charge both enlarge the maximal black hole size compatible with a de Sitter background.

\subsubsection*{Ultracold Black Hole}

The ultracold configuration arises when all three horizons---the cosmological, event, and Cauchy horizons---coincide:
\begin{equation}
    r_{C} = r_{H} = r_{C_H} \equiv r_{\text{uc}} \, .
\end{equation}
This represents the maximal allowed horizon degeneracy. The resulting near-horizon geometry becomes highly symmetric and can be written as a nontrivial fibration
\begin{equation}
    \mathbb{M}^{1,1} \times_{f} S^{1} ,
\end{equation}
which, via an appropriate coordinate transformation \cite{Booth:1998gf}, may also be expressed as a fibration of two-dimensional Rindler space over a circle. Both descriptions highlight the thermodynamically extreme nature of the ultracold limit, where
\begin{equation}
    T_{\text{uc}} = 0 \, ,
\end{equation}
and the entropy reaches its minimal value.

In the non-rotating limit the ultracold branch ceases to exist; instead, the solution approaches the static Nariai configuration. Thus, rotation is essential for supporting a fully degenerate ultracold horizon.

\subsubsection*{Lukewarm Case}

The quantum Kerr--dS$_3$ geometry also admits a lukewarm branch, defined by equality of the surface gravities of the event and cosmological horizons,
\begin{equation}
    \kappa_{H} = \kappa_{C} \, .
\end{equation}
Despite this thermodynamic equilibrium, the two horizons remain distinct. The lukewarm regime is therefore non-extremal and contrasts with the Nariai and ultracold limits, where horizon coincidence occurs.

Physically, the lukewarm solution describes a thermal balance between the black hole and the surrounding de Sitter region. Because both horizons share a common temperature, semiclassical field theory on this background remains well-defined, free from the divergences or instabilities associated with degenerate or widely separated horizons.

\begin{figure}[h!]
\centering
\includegraphics[width=0.7\textwidth]{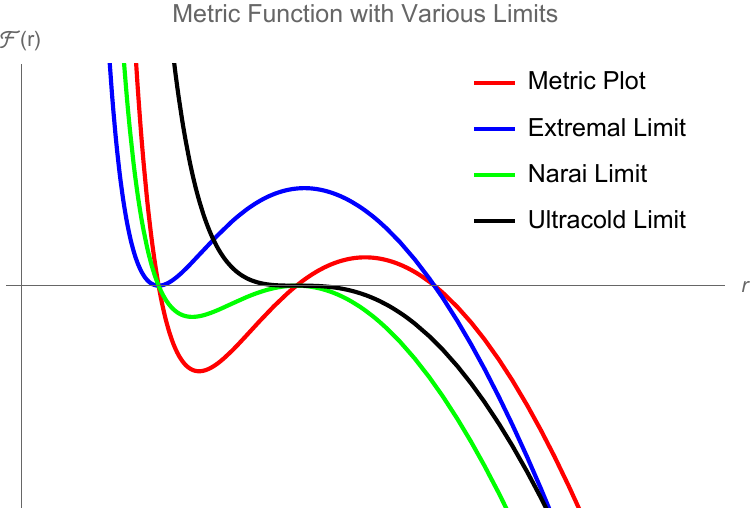}
\caption{Qualitative behaviour of the metric function $\mathcal{F}(r)$, illustrating the various horizon configurations of the quantum Kerr--dS$_3$ spacetime.}
\label{fig:Metric function Plot}
\end{figure}

The nature of the horizon structure is governed by the sign of $g_{\bar{r}\bar{r}}$ evaluated at its minimum, $r=r_{\min}$. If $g_{\bar{r}\bar{r}}(r_{\min}) < 0$, the spacetime supports multiple non-degenerate horizons. When $g_{\bar{r}\bar{r}}(r_{\min}) = 0$, a double root appears, indicating an extremal configuration. If $g_{\bar{r}\bar{r}}(r_{\min}) > 0$, no real positive root exists, and the geometry exhibits a naked singularity.

\medskip

The four-dimensional bulk entropy follows from the Bekenstein--Hawking area law,
\begin{equation}
S^{(4)}_{\mathrm{BH}}
= \frac{\pi}{G_{4}}
\, \frac{\eta \ell y_{1}\left(r_{i}^{2} + a^{2} y_{1}^{2}\right)}{\ell + r_{i} y_{1}} \, .
\end{equation}
In the $\alpha=0$ limit these expressions reduce to the known thermodynamics of the static $\mathrm{AdS}_{4}$ bulk \cite{Emparan:2022ijy}. Equivalent thermodynamic relations follow from a canonical-ensemble evaluation of the on-shell bulk action, paralleling the method of Ref.~\cite{Kudoh:2004ub}. The first law takes the form
\begin{equation}
    dM = T_{i}\, dS^{(4)}_{\mathrm{BH}} + \Omega_{i}\, dJ \, ,
\end{equation}
valid for arbitrary parameter values, including variations of the brane tension encoded by $\nu$.

\subsection{The Extremal Solution}

The extremal horizon is located at the point where the event horizon coincides with the minimum of the metric function $g_{\bar{r}\bar{r}}(\bar{r})$, satisfying
\begin{equation}\label{Extrimality condition_in_metric_function}
    g_{\bar{r}\bar{r}}(r_{H}) = 0 \,, 
    \qquad 
    \partial_{r} g_{\bar{r}\bar{r}}(r)\big|_{r=r_{H}} = 0 \, .
\end{equation}

Since $g_{\bar{r}\bar{r}}(\bar r)$ contains inverse powers of $\bar r$, it is convenient to define
\begin{equation}
    \mathcal{Q}(\bar r) \equiv \bar r^{2} g_{\bar{r}\bar{r}}(\bar r) \, ,
\end{equation}
which is polynomial in $\bar r$. Extremality is then equivalent to
\begin{equation}
    \mathcal{Q}(r_{H}) = 0 \,, \qquad \mathcal{Q}'(r_{H}) = 0 \, .
\end{equation}

Solving these yields
\begin{align}\label{E_cond_in_M_J}
r_{H} &= \frac{\sqrt{-\mathscr{F} + \ell_{3}^{2}(1 - 8\mathcal{G}_{3} M) + 4 r_{S}^{2}}}{\sqrt{6}} \ , \nonumber \\[2mm]
\mu &= 
\frac{\sqrt{\frac{2}{3}}
\left(\mathscr{F} + \ell_{3}^{2}(2 - 16\mathcal{G}_{3} M) - 4r_{S}^{2}\right)
\sqrt{-\mathscr{F} + \ell_{3}^{2}(1 - 8\mathcal{G}_{3} M) - 2 r_{S}^{2}}
}
{3(\tilde a^{2}+1)^{3/2} \Delta^{3} \ell\, \ell_{3}^{2}} ,
\end{align}
where
\begin{equation}
\mathscr{F}
= \sqrt{
64 \mathcal{G}_{3}^{2} \ell_{3}^{2} (\ell_{3}^{2}M^{2}-3J^{2})
-16 \mathcal{G}_{3} \ell_{3}^{2} M (\ell_{3}^{2}-8 r_{S}^{2})
+ \ell_{3}^{4} - 16\ell_{3}^{2}r_{S}^{2} + 16 r_{S}^{4}
} \, .
\end{equation}
Thus, the extremal horizon radius $r_{H}$ and the corresponding parameter $\mu$ are completely determined.

\section{Weak Cosmic Censorship Conjecture}\label{Sec:Test of wCCC}

In this section, we begin by analyzing the dynamical behavior of test particles in the background of the qKdS$_3$ black hole. In particular, we consider geodesic trajectories of infalling particles and examine the conditions under which they can reach the event horizon. Owing to the presence of a strong centrifugal barrier, particles with sufficiently large angular momentum are prevented from being absorbed by the black hole. Thus, identifying the maximum allowed angular momentum for particle capture is essential to delineate the range of dynamically admissible perturbations. Following the dynamical analysis, we turn to the perturbative framework required to test the wCCC. Specifically, we derive the first- and second-order perturbative inequalities associated with matter field interactions in the qKdS$_3$ background. These inequalities establish the necessary conditions under which an extremal black hole might be pushed beyond extremality, potentially leading to a naked singularity. The combination of the geodesic analysis and perturbative approach allows for a comprehensive examination of the wCCC in the qKdS$_3$ black hole.

Now, we rewrite the Eq.~\eqref{E_cond_in_M_J} in the terms of $y_1$ and $\Tilde{a}$ as 
\begin{eqnarray}
    r_H &=& \sqrt{\frac{2}{3}} \ell_3 \sqrt{\frac{\Tilde{a}^2 \left(12-5 y_1^2\right)+y_1^2}{\left(\Tilde{a}^2-y_1^2+3\right)^2}-\sqrt{\frac{\mathscr{F_A}}{\left(\Tilde{a}^2-y_1^2+3\right)^4}}} \nonumber \\ 
    \mu \nu &=& \frac{\sqrt{\frac{2}{3}} \left(2 \left(\Tilde{a}^2+1\right) y_1^2+\mathscr{F_A}\right) \sqrt{ \left(\left(\Tilde{a}^2+1\right) y_1^2-\mathscr{F_A}\right)}}{3 \left(\Tilde{a}^2+1\right)^{3/2}  y_1^3}
\end{eqnarray}
where 
\begin{eqnarray}
    \mathscr{F_A} = \sqrt{\left(\Tilde{a}^2+1\right)^2 y_1^4-48 \Tilde{a}^2 y_1^2-12 \left(\Tilde{a}^6+6 \Tilde{a}^4+\Tilde{a}^2\right)} \ .
\end{eqnarray}
Here, we introduce a convenient dimensionless parameter that encodes the effect of the backreaction, defined as
$$
\nu \equiv \frac{\ell}{\ell_{3}} \ .
$$
Here, $\ell$ characterizes the strength of the backreaction.  The ratio $\nu$ thus measures the relative scale of the backreaction with respect to the AdS background, and it is straightforward to see that the zero-backreaction limit can be obtained by $\nu \;\longrightarrow\; 0$, which recovers the unperturbed AdS$_3$ geometry. Conversely, finite values of $\nu$ correspond to non-trivial backreaction effects that deform the extremality condition and the thermodynamic behavior of the black hole.

\subsection{Dynamics of test particle}

In this subsection, we examine the dynamics of test particles in the given spacetime. The trajectories of these particles are determined by the geodesic equation, which describes how a particle moves by extremizing the proper time in a curved geometry. This reflects the principle that free-falling particles follow the straightest possible paths, or geodesics, under the influence of gravity alone. Alternatively, the particle motion can be analyzed using the Lagrangian formulation. In this approach, the motion is derived from a Lagrangian constructed using the spacetime metric and the particle’s velocity. This method is particularly useful for identifying conserved quantities arising from symmetries of the spacetime. The spacetime is stationary and axially symmetric, meaning it possesses symmetries under time translations and rotations around an axis. These symmetries imply the existence of conserved quantities due to Noether’s theorem. Specifically, the symmetry under time translation leads to the conservation of the particle's energy per unit mass, while the rotational symmetry implies conservation of the angular momentum per unit mass.

The conserved quantities, often denoted by $\delta\,\mathrm{e}$ for energy and $\delta\,\mathrm{L}$ for angular momentum, are associated with the spacetime’s Killing vectors. They play a crucial role in determining the allowed motion of test particles, especially in relation to black hole horizons and potential barriers in the geometry. Following \cite{Frassino:2024fin}, form of these conserved quantities is
\begin{eqnarray}\label{conjugatemomenta}
\delta\,\mathrm{e} = - g_{tt} \frac{dt}{d\tau} - g_{t\varphi} \frac{d\varphi}{d\tau} \;\;\;\;\;\text{and}\;\;\;\;\; \delta\,\mathrm{L} = g_{\varphi\varphi} \frac{d\varphi}{d\tau} + g_{t\varphi} \frac{dt}{d\tau} \ .
\end{eqnarray}
Solving Eqs.~\eqref{conjugatemomenta} for the time and angular components of the four-velocity yields
\begin{eqnarray}\label{tdot}
\frac{dt}{d\tau} = \frac{g_{t\varphi} \delta\,\mathrm{L} + g_{\varphi\varphi} \delta\,\mathrm{e}}{g_{t\varphi}^2 - g_{tt} g_{\varphi\varphi}} \;\;\;\;\;;\;\;\;\;\; \frac{d\varphi}{d\tau} = -\frac{g_{tt} \delta\,\mathrm{L} + g_{t\varphi} \delta\,\mathrm{e}}{g_{t\varphi}^2 - g_{tt} g_{\varphi\varphi}}\,. \label{phidot}
\label{dottdotphi}
\end{eqnarray}
When a test particle crosses the event horizon and is absorbed by the black hole, it ceases to exist as an independent entity and becomes part of the black hole. As a result, the conserved quantities associated with the spacetime, namely the ADM mass \( M \) and total angular momentum \( L \), are modified due to the energy \( \mathrm{e} \) and angular momentum \( \mathrm{L} \) carried by the particle. The transition to the new configuration is described by
\[
M \rightarrow M' = M + \mathrm{e}, \quad L \rightarrow L' = L + \mathrm{L}.
\]
This process effectively transforms the original black hole into a new stationary solution characterized by the updated mass and angular momentum, thereby forming a composite gravitational system. When a test particle is located outside the event horizon of the black hole, its motion is characterized by a timelike four-velocity, consistent with the requirements of a physical trajectory in general relativity. This implies that the particle follows a path with positive proper time and causal structure appropriate for the particles. By analyzing the spacetime geometry and the conserved quantities associated with the particle’s motion, such as energy and angular momentum, one can determine how the black hole’s mass energy changes upon absorption of the particle. Specifically, it is found that the energy imparted to the black hole by the test particle can, in principle, take two distinct values. However, a physically meaningful solution must correspond to a timelike, future-directed trajectory. This imposes a constraint on the motion—namely, that the time component of the particle's four-velocity remains positive throughout its infall. This condition ensures that the particle moves forward in coordinate time as it approaches and eventually crosses the event horizon in alignment with the causal structure of the black hole spacetime. By examining the conserved quantities associated with the test particle—namely its energy and angular momentum—and imposing the necessary requirements for physically viable motion (i.e., that the trajectory remains timelike and future-directed), one arrives at specific constraints these quantities must satisfy. To determine the maximal value of the angular momentum \( L \) for which the black hole can absorb a test particle, we adopt the criterion of~\cite{Wald:1974hkz}. Specifically, we require that the particle's geodesic trajectory be future-directed, which in a stationary spacetime translates to the condition \( \dot{t} > 0 \). Applying this to Eq.~\eqref{tdot}, we obtain the upper bound on the particle’s angular momentum in terms of its energy and the background metric components
\begin{equation}
L < -\frac{g_{\varphi\varphi}}{g_{t\varphi}} E \ .
\label{Lbound}
\end{equation}
This condition delineates the region of phase space where a particle possesses sufficiently low angular momentum to overcome the centrifugal barrier and cross the event horizon. Particles with angular momentum exceeding this threshold are prevented from reaching the horizon due to repulsive centrifugal effects and are instead reflected back to infinity. Thus, Eq.~\eqref{Lbound} provides a necessary condition for horizon penetration and is essential in assessing the potential for extremality violation in the context of the weak cosmic censorship conjecture. To ensure that a particle is absorbed by the black hole rather than scattered or reflected, the condition \eqref{Lbound} must be upheld across the entire domain exterior to the event horizon, i.e., for all \( r \geq r_H \). From a dynamical perspective, when the test particle possesses sufficiently large angular momentum, the centrifugal barrier becomes dominant over the gravitational attraction exerted by the black hole. As a result, the particle is unable to reach the event horizon and instead follows a trajectory that asymptotically deviates from the black hole. In other words, the black hole cannot absorb a test particle with angular momentum exceeding a critical threshold. Therefore, for the particle to be captured, its angular momentum must remain below a maximum limit, denoted by \( \mathrm{L}_{\text{max}} \). This upper limit on \( \mathrm{L} \) can be determined explicitly as
\begin{equation}\label{Lmax}
\mathrm{L}_\text{max} = \frac{y_1}{4\mathcal{G}_3
}\left[\frac{\bar{r}\sqrt{\bar{r}^2-r_s^2}+\mu \ell \sqrt{\tilde{a}^2+\tilde{a}^4}\,\Delta^2 R_3^2}{y_1 \sqrt{\bar{r}^2-r_s^2} + \ell \Delta \sqrt{1+\tilde{a}^2}} \;\mathrm{e}\right]_{\bar{r}=\bar{r}_H}
\end{equation}
This is a function of both the energy \( \mathrm{e} \) of the infalling particle and the black hole background parameters, i.e., \( (\tilde{a}, y_1, \nu, \ell_3, \kappa) \), also \( L_\text{max} > 0 \) over the domain of parameters considered in this work. This positivity ensures the existence of infalling particles that satisfy the necessary kinematic condition for potential over-spinning. 
This analysis indicates that an upper bound exists on the angular momentum of an infalling test particle.


\subsection{Perturbation Analysis}\label{ssec_order1}

We consider the scenario in which an extremal black hole absorbs a particle characterized by energy \( \mathrm{e} \) and angular momentum $\mathrm{L}_{max}$ as in Eq.~\eqref{Lmax}. We then investigate the possible final configurations that the system may evolve into following the absorption process. Assuming that the final state is also a stationary geometry belonging to the same class of qKdS$_3$ spacetimes. In the extremal configuration, the metric function \( g_{{\bar{r}\bar{r}}} \) attains its global minimum precisely at the event horizon radius, ensuring that the horizon is degenerate. Upon the absorption of a particle, the background geometry is perturbed, resulting in a shift of the minimum of \( g_{{\bar{r}\bar{r}}} \) to a new location, denoted \( r_\text{min} \), which may not correspond to a valid event horizon. A robust criterion for diagnosing the validity of the weak cosmic censorship conjecture in the perturbed geometry is to evaluate the sign of the metric function at the post-absorption minimum \( g_{{\bar{r}\bar{r}}}(r_\text{min}) \). 

To analyze the stability of the extremal black hole solution under infalling matter, we adopt a perturbative approach and focus on the linearized response of the background geometry following the absorption of a test particle carrying the maximal allowed angular momentum. In this framework, the black hole parameters undergo infinitesimal shifts \( \tilde{a} \to \tilde{a} + \delta \tilde{a} \) and \( y_1 \to y_1 + \delta y_1 \), inducing a corresponding displacement in the radial position of the minimum of the metric function \( g_{{\bar{r}\bar{r}}} \), denoted by \( r_\text{min} \to r_\text{min} + \delta r \). The value of the metric function evaluated at the perturbed minimum can then be expanded to first order as
\begin{equation}
\mathcal{F} (\bar{r}_\text{min} + \delta \bar{r},\, \tilde{a} + \delta \tilde{a},\, y_1 + \delta y_1) = \mathcal{F}(\bar{r}_\text{min},\, \tilde{a},\, y_1) + \delta \mathcal{F} + \mathcal{O}(\delta^2 \mathcal{F}) \ ,
\label{Hexpansiongeneral}
\end{equation}
where the linear correction \( \delta \mathcal{F} \) is 
\begin{equation}\label{Variation of grr}
\delta \mathcal{F} = \left. \frac{\partial \mathcal{F}}{\partial \bar{r}} \right|_{\bar{r} = \bar{r}_\text{min}} \delta \bar{r} 
+ \left. \frac{\partial \mathcal{F}}{\partial \tilde{a}} \right|_{\bar{r} = \bar{r}_\text{min}} \delta \tilde{a} 
+ \left. \frac{\partial \mathcal{F}}{\partial y_1} \right|_{\bar{r} = \bar{r}_\text{min}} \delta y_1 \ .
\end{equation}
All partial derivatives are understood to be evaluated on the unperturbed extremal background. At leading order in perturbation theory, the variations \( \delta \tilde{a} \) and \( \delta y_1 \) induced by the absorption of a test particle can be systematically related to the conserved quantities of the particle, namely its energy \( E \) and angular momentum \( L_\text{max} \). Upon absorption, the black hole’s global parameters undergo the following infinitesimal shifts
\begin{equation}
\begin{split}
M &\to M + \delta M \;\;\;\;\;\;\;\;\;;\;\;\;\;\;\;\;\;\; J \to J + \delta J \ ,
\end{split}
\label{MJchange}
\end{equation}
where \( \delta M \) and \( \delta J \) denote the respective perturbations in the ADM mass and total angular momentum. Since $M$ and $J$ are the functions of $\tilde{a}$ and $y_1$ then these variations can be expressed in terms of variations in the intrinsic black hole parameters \( \tilde{a} \) and \( y_1 \) as
\begin{subequations}
\begin{align}
\delta M = \left.\frac{\partial M}{\partial \tilde{a}}\right|_{\text{ext}} \delta \tilde{a} + \left.\frac{\partial M}{\partial y_1}\right|_{\text{ext}} \delta y_1 \;\;\;\;\;\;\;\;\text{and}\;\;\;\;\;\;\;\;\; \delta J = \left.\frac{\partial J}{\partial \tilde{a}}\right|_{\text{ext}} \delta \tilde{a} + \left.\frac{\partial J}{\partial y_1}\right|_{\text{ext}} \delta y_1\,.
\end{align}
\label{deltaMdeltaJ}
\end{subequations}
The sign of \( \delta \mathcal{F} \) thus encodes the leading-order response of the near-horizon geometry to the perturbation and serves as a diagnostic tool for testing the weak cosmic censorship conjecture. In particular, \( \delta \mathcal{F} > 0 \) implies that the metric function becomes strictly positive at its minimum, and no real root exists for \( \mathcal{F} = 0 \), signaling the formation of a naked singularity. Conversely, \( \delta \mathcal{F} \leq 0 \) ensures the persistence of an event horizon, either non-degenerate or extremal, consistent with the preservation of cosmic censorship. Let's compute \( \delta \mathcal{F} \) using Eq.~\eqref{qKdS Metric} as
\begin{eqnarray}
   \mathcal{F}(\bar{r},\tilde{a},y_1) = 1 - 8\mathcal{G}_{3}M(\tilde{a},y_1) - \frac{\bar{r}^{2}}{\ell_{3}^{2}} + \frac{(4\mathcal{G}_{3}J(\tilde{a},y_1))^{2}}{\bar{r}^{2}} - \Psi(\bar{r},\tilde{a},y_1) \ .
\end{eqnarray}
Now, using the Eq.~\eqref{Variation of grr}, we have
\begin{eqnarray}\label{Variation of grr2}
    \delta \mathcal{F} &=& \left. \left( -2 \frac{\bar{r}}{\ell_{3}^{2}} -2 \frac{(4\mathcal{G}_{3}J(\tilde{a},y_1))^{2}}{\bar{r}^{3}} - \partial_{\bar{r}} \Psi(\bar{r},\tilde{a},y_1)\right)\right|_{\bar{r} = \bar{r}_\text{min}} \delta \bar{r} \nonumber \\
&& + \left. \left(- 8\mathcal{G}_{3} \partial_{\ta}M(\tilde{a},y_1) + \frac{32\mathcal{G}_{3}^2 J(\tilde{a},y_1)}{\bar{r}^{2}} \partial_{\ta} J(\tilde{a},y_1)- \partial_{\ta}\Psi(\bar{r},\tilde{a},y_1)\right) \right|_{r = r_\text{min}} \delta \tilde{a}   \\
&& + \left. \left(- 8\mathcal{G}_{3} \partial_{y_1}M(\tilde{a},y_1) + \frac{32\mathcal{G}_{3}^2 J(\tilde{a},y_1)}{\bar{r}^{2}} \partial_{y_1} J(\tilde{a},y_1)- \partial_{y_1}\Psi(\bar{r},\tilde{a},y_1)\right)  \right|_{r = r_\text{min}} \delta y_1 \ . \nonumber
\end{eqnarray}
Given that the unperturbed solution is an extremal black hole, the condition $\mathcal{F} |_{r=r_{min}}=0$ at the horizon $r_{min}$ gives us
\begin{equation}\label{Psi at rh}
    \Psi(\bar{r}_H,\tilde{a},y_1) = 1 - 8\mathcal{G}_{3}M - \frac{\bar{r}_H^{2}}{\ell_{3}^{2}} + \frac{(4\mathcal{G}_{3}J)^{2}}{\bar{r}_H^{2}} \ .
\end{equation}
Given that $\partial_{\ta}\Psi(\bar{r},\tilde{a},y_1) |_{r = r_{min}}= \partial_{\ta}\Psi(\bar{r}_{min},\tilde{a},y_1)$ (and similarly for the parameter $y_1$), it is easy to show that 
\begin{eqnarray}\label{Psi Variation at rh}
    \partial_{\ta}\Psi(\bar{r}_H, \tilde{a}, y_1) &=&  - 8\mathcal{G}_{3}\partial_{\ta} M  + \frac{32\,\mathcal{G}_{3}^2 J}{\bar{r}_H^{2}}\partial_{\ta}J \\
    \partial_{y_1}\Psi(\bar{r}_H, \tilde{a}, y_1) &=&   - 8\mathcal{G}_{3}\partial_{y_1} M  + \frac{32\,\mathcal{G}_{3}^2 J}{\bar{r}_H^{2}}\partial_{y_1}J \ .\nonumber
\end{eqnarray}
Finally, putting Eq.\eqref{Psi Variation at rh} in Eq.\eqref{Variation of grr2} and using the fact that $\partial _{\bar{r}} \mathcal{F}|_{\bar{r}=\bar{r}_{min}}=0$ for extremal solutions, we can conclude 
 \begin{equation}
     \delta \, \mathcal{F} = 0 \ .
 \end{equation}

In the Wald–Sorce framework, the sign of the perturbed metric function at its minimum determines whether the horizon persists after the absorption process. A negative or vanishing first-order variation, $\delta \mathcal{F}\leq 0$, signals that the perturbed configuration continues to admit at least one real root of $ \mathcal{F}(r)$, and therefore retains an event horizon. By contrast, $\delta \mathcal{F} > 0$ would imply that the minimum of $\mathcal{F}(r)$ becomes strictly positive, eliminating all real roots and exposing the central singularity. In our analysis the extremal $qKdS_3$ background satisfies $\delta \mathcal{F} = 0$, which means that the degenerate zero of the metric function is preserved under the perturbation. Thus the horizon neither disappears nor splits, and the black hole remains extremal. This equality therefore represents successful preservation of cosmic censorship rather than a marginal or ambiguous case.

For the extremal black holes, the metric function $\mathcal{F}(r)$ attains its global minimum at the degenerate event horizon, where the inner and outer horizons coalesce. This configuration corresponds to a zero-temperature black hole with vanishing surface gravity. When a scattering process occurs, it is found that the minimum of $\mathcal{F}(r)$ remains unaltered. This invariance implies that the essential geometric and causal structure—specifically, the presence of a degenerate horizon—remains intact throughout the interaction. The preservation of the horizon’s degeneracy indicates that the extremal nature of the black hole is dynamically stable under such perturbations. As a result, the black hole does not transition to a non-extremal configuration, which would otherwise involve a separation of the horizons and the emergence of a nonzero temperature. The persistence of extremality ensures that no naked singularities are exposed, thereby maintaining the causal integrity of the spacetime. Since the scattering process does not induce a breakdown of the event horizon, the central singularity remains enclosed, preventing any violation of predictability. The constancy of the minimum of $\mathcal{F}(r)$ thus serves as a sufficient condition to affirm that extremal black holes retain their horizon structure under such processes, reinforcing the validity of the wCCC in this context.


We can go beyond the first order and analyze the second-order correction to the metric function \(\mathcal{F}\), induced by perturbations of the parameters \(\tilde{a} \to \tilde{a} + \delta \tilde{a}\), \(y_{1} \to y_{1} + \delta y_{1}\), and \(r_{\min} \to r_{\min} + \delta r\).\footnote{Several works in the literature (see, e.g., \cite{Sorce:2017dst, Ding:2020zgg}) implement second--order perturbations in the mass and angular momentum, \((\delta M + \delta^{2}M)\) and \((\delta J + \delta^{2}J)\), which is equivalent to introducing second-order variations in \(\tilde{a}\) and \(y_{1}\). In the present context such an expansion is unnecessary: from the viewpoint of the black hole, the absorbed particle modifies the ADM charges only through the combinations \(\delta \tilde{M}\) and \(\delta \tilde{J}\), rendering the explicit inclusion of \(\delta^{2}\tilde{a}\) and \(\delta^{2}y_{1}\) superfluous.}
The second-order variation of the metric function is obtained as
\begin{align}
\delta^{2} \mathcal{F} = &\frac{1}{2}\left[\left.\partial_{\bar r}^{2}\mathcal{F}\right|_{\bar r=\bar r_{\min}} (\delta\bar r)^{2} + \left.\partial_{\tilde a}^{2}\mathcal{F}\right|_{\bar r=\bar r_{\min}} (\delta\tilde a)^{2} + \left.\partial_{y_{1}}^{2}\mathcal{F}\right|_{\bar r=\bar r_{\min}} (\delta y_{1})^{2} \right] \notag \\ 
&\quad + \left.\partial_{\tilde a}\partial_{y_{1}}\mathcal{F}\right|_{\bar r=\bar r_{\min}}  \, \delta\tilde a\,\delta y_{1} + \left.\partial_{\tilde a}\partial_{\bar r}\mathcal{F}\right|_{\bar r=\bar r_{\min}} \, \delta\tilde a\,\delta\bar r + \left.\partial_{\bar r}\partial_{y_{1}}\mathcal{F}\right|_{\bar r=\bar r_{\min}} \, \delta\bar r\,\delta y_{1}.
\label{eq:delta2F}
\end{align}

As shown in Sec.~\ref{ssec_order1}, once the extremality conditions are imposed, all mixed and parameter--derivative terms vanish, and only the term 
\(\partial_{\bar r}^{2}\mathcal{F}|_{\bar r=\bar r_{\min}}\) survives.  
Analytically, this remaining contribution is strictly positive for a finite interval of \(\tilde a\), reflecting the fact that \(r_{\min}\) corresponds to a local minimum of \(\mathcal{F}\).  
This behaviour is illustrated in Fig.~\ref{fig:Metric function Plot}.\footnote{
To render symbolic manipulations tractable, we set 
\(\mathcal{G}_{3}=1\), \(\ell_{3}=1\), and \(\ell=1\) in Fig.~\ref{fig:Metric function Plot}. 
This restriction is technical and does not affect the physical conclusions.}

Although second--order corrections yield a nonzero quantity \(\delta^{2}\mathcal{F}\), the physical interpretation requires care.  
The purpose of the perturbation analysis is to determine whether the absorption of a particle carrying mass \(\delta M\) and angular momentum \(\delta J\) can evolve an extremal qKdS black hole into a configuration that violates the weak cosmic censorship conjecture.  
Mathematically, this corresponds to identifying all contributions to \(\mathcal{F}\) proportional to 
\(\delta M\), \(\delta J\), \((\delta M)^{2}\), and \((\delta J)^{2}\).  
We have shown that all such terms vanish in the extremal limit, and hence no violation of the wCCC is possible at second order.  
The remaining terms in \(\delta\mathcal{F}\) and \(\delta^{2}\mathcal{F}\) arise solely from the intrinsic extremal geometry via \(\delta\bar r\) and \((\delta\bar r)^{2}\), and therefore do not correspond to dynamical perturbations of the black hole parameters.
\begin{figure}[t]
\centering
\includegraphics[width=0.7\textwidth]{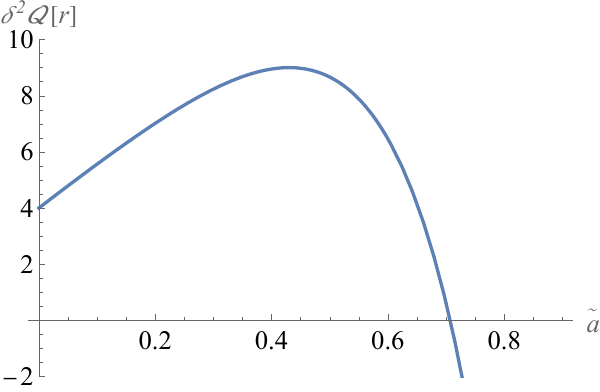}
\caption{Second--order variation \(\delta^{2}\mathcal{Q}\) vs $\Tilde{a}$. Here $\mathcal{Q}(r) =r^2 \mathcal{F}(r)$.}
\label{fig:Metric function Plot}
\end{figure}
A fully general second-order analysis may be carried out numerically. We do not reproduce the full computational procedure here, since it has been extensively discussed in the literature. The essential strategy is to solve the extremality conditions numerically for the admissible branch, perturb the parameters in terms of \(\delta M\) and \(\delta J\), and expand the resulting expressions to quadratic order. One then evaluates \(\min g^{rr}(r)\) to extract the second-order behaviour of the effective potential, reproducing the qualitative structure exhibited in~\cite{Frassino:2024fin, Amo:2025phg}.

\section{Conclusions}\label{sec:conclusions}

In this work, we have extended Wald’s seminal gedanken experiment to the quantum Kerr--de~Sitter black hole in three dimensions (qKdS$_3$), a geometry that incorporates the exact one-loop backreaction of strongly coupled conformal matter. Our analysis has focused on the interaction between finely tuned test particles and extremal qKdS$_3$ black holes, with the objective of assessing the robustness of their horizons under perturbations and, consequently, the validity of the Weak Cosmic Censorship Conjecture (wCCC) in a setting where quantum corrections are fully incorporated. We find that quantum backreaction does not modify the qualitative outcome of Wald-type extremality tests: test particles fail to destroy the degenerate horizon, and the quantum-corrected geometry remains consistent with the wCCC.

A central result of this paper is the extension of the analysis beyond linear order in perturbations. While the first-order test already indicates that extremal qKdS$_3$ black holes cannot be overspun, we have performed a second-order perturbative study of the metric function~$\mathcal{F}$. By examining variations of the rotation parameter, the auxiliary parameter $y_1$, and the minimum-radius condition, we have shown that both the first- and second-order corrections vanish when evaluated at the extremal point where the inner and outer horizons coincide. Physically, this implies that the scattering process leaves the minimal value of the metric function invariant. Since extremality is encoded in the presence of a degenerate zero of $\mathcal{F}$, the invariance of this minimum ensures that horizon degeneracy is preserved under the perturbation, and the black hole is prevented from transitioning to a super-extremal configuration. Thus, quantum backreaction, while modifying the background geometry in a nontrivial manner, does not destabilize the extremal horizon.

An additional outcome of our study is methodological: it establishes that the second-order wCCC test can be successfully implemented for qKdS$_3$ spacetimes, even when analytic simplifications are restricted to the tractable case $\mathcal{G}_{3}=\ell_{3}=\ell=1$. Although this simplification facilitates symbolic manipulation, it does not alter the underlying physics. A complete analysis across the full parameter space---including arbitrary values of $\mathcal{G}_{3}$, $\ell_{3}$, and $\ell$---is expected to require numerical techniques.

In summary, our results provide strong evidence that the Weak Cosmic Censorship Conjecture remains valid for qKdS$_3$ black holes, even when quantum backreaction and second-order perturbations are taken into account. Nevertheless, several avenues for further research remain open. An important extension would be to perform the gedanken experiment in a \emph{near-extremal} rather than exactly extremal background. Near-extremal configurations often offer a more sensitive probe of possible horizon instabilities and may reveal whether quantum corrections inhibit or enhance the prospects for overspinning in regimes that approach, but do not reach, extremality. Such investigations may shed additional light on the interplay between quantum effects and horizon stability and may contribute to a deeper understanding of how predictability is preserved in quantum-corrected black hole spacetimes.

\section*{Acknowledgement }

We gratefully acknowledge Antonia M. Frassino for insightful and stimulating discussions that significantly contributed to the development of this work. Ankit Anand is financially supported by the Institute's postdoctoral fellowship at IIT Kanpur.

\appendix


\section*{Data Availability Statement}
Data sharing is not applicable to this article, as no datasets were generated or analyzed during the current study.



\bibliographystyle{JHEP}

\bibliography{biblio}
\end{document}